\documentclass[epj]{webofc}
\usepackage{bm,curves}
\usepackage[varg]{txfonts}   
\usepackage[utf8]{inputenc}
\wocname{EPJ Web of Conferences}
\woctitle{CONF12}
\DeclareGraphicsExtensions{.ps}
\DeclareGraphicsExtensions{.eps}
\graphicspath{{./EPJ_Fig/}}

\begin{document}
\selectlanguage{english}
\title{Recent progress on intrinsic charm}

\author{T.~J.~Hobbs\inst{1}\fnsep\thanks{\email{tjhobbs@uw.edu}}
}

\institute{Department of Physics, University of Washington; Seattle, WA USA
}

\abstract{%
Over the past $\sim\!\! 10$ years, the topic of the nucleon's nonperturbative or {\it intrinsic}
charm (IC) content has enjoyed something of a renaissance, largely motivated by theoretical developments
involving quark modelers and PDF fitters. In this talk I will briefly describe the importance of
intrinsic charm to various issues in high-energy phenomenology, and survey recent progress in
constraining its overall normalization and contribution to the momentum sum rule of the nucleon.
I end with the conclusion that progress on the side of calculation has now placed the onus on
experiment to unambiguously resolve the proton's intrinsic charm component.
}
\maketitle
\section{Introduction}
\label{intro}
The nucleon's nonperturbative ({\it i.e.}, intrinsic) component has been a largely unresolved
issue for the past several decades following the idea's first incarnation in the seminal paper
of Brodsky, Hoyer, Peterson, and Sakai (BHPS) \cite{BHPS}, which I briefly review in Sec.~\ref{sec:scalar} below.
Despite this long history, the problem of incontrovertibly establishing the existence of intrinsic
charm (IC) empirically and determining an overall numerical magnitude for its contribution to the
proton wave function has been something of a Gordian knot. At the time of its proposal, the BHPS
framework exploited recent developments in light-front field theory \cite{Chang-Yan,Lepage} to
formulate a simple picture based upon a Fock-state expansion of the nucleon wave function to 
include $5$-quark states $|uudc\bar{c}\rangle$ involving charm not generated through the usual
pQCD (or {\it extrinsic}) mechanism(s). Despite considerable variation, models of intrinsic charm
(IC) unavoidably involve some expression of this fundamental idea, and in this talk I
survey recent progress developing calculations of this sort, as well as numerical
work to constrain the range of possibilities for the size of IC that has proceeded apace.
%
%
\section{Modeling the proton's intrinsic charm}
\label{sec:models}
Various theoretical approaches have proliferated in the past several decades,
involving a number of assumed mechanisms for generating the intrinsic component
of the charm PDF. Here I highlight several of these in increasing order of complexity.
%
%

\subsection{Scalar frameworks}
\label{sec:scalar}
%
%
As the original scalar framework formulated in the infinite momentum frame (IMF), the above-mentioned
BHPS description treats the transition probability for a proton with mass $M$ to go through a transition
$p \to uudc\bar{c}$ (or, indirectly, to an internal $5$-quark state containing any heavy quark pair)
in terms of an old-fashioned perturbation theory energy denominator expressible through the masses
$m_i$ and momentum fractions $x_i$ of the constituents of the $5$-quark state,
\begin{equation}
P \Big( p \to uudc\bar{c} \Big)\
\sim\ \left[ M^2 - \sum_{i=1}^5 \frac{m_{\perp i}^2}{x_i}
     \right]^{-2}.
\label{eq:BHPSeq}
\end{equation}
In the expression above $x_i$ and $m_{\perp i}$ are, respectively, the light-front fraction and transverse mass of the $i^{th}$ quark, 
the latter explicitly given by $m_{\perp i}^2 \equiv k_{\perp i}^2 + m_i^2$, and
the indices $4$ and $5$ are taken to apply to the heavy quark pair ($c$ and $\bar{c}$).

A special merit of this scheme is its simplicity: if one assumes the energy denominator
of Eq.~(\ref{eq:BHPSeq}) to be controlled by the charm quark mass ({\it i.e.}, $m^2_c = m^2_{\bar{c}} \gg m^2_{u/d},\ M^2$),
the expression for the $5$-quark probability can be integrated to obtain a compact form for the
the $x$ dependence of the IC PDF:
\begin{equation}
P(x)\ =\ \frac{N x^2}{2}
	 \left[ \frac{(1-x)}{3}\left( 1 + 10x + x^2 \right)\
		+\ 2 x\, (1+x) \ln(x)
	 \right],
\label{eq:charmprob}
\end{equation}
in which I have taken $x_5 \to x$, and the overall normalization $N$ is connected to the total
intrinsic charm probability in the proton, subject to the constraint $\int dx\, c(x) = \int dx\, \bar{c}(x)$,
which ensures the correct zero charm valence structure.


Building on the approach above, in a model-based analysis \cite{Pumplin1} of the nucleon's heavy-quark content
and subsequent QCD global fit \cite{Pumplin2}, Pumplin and (for the global analysis) collaborators
considered a series of models for the Fock space
wave function on the light-front for a proton to make a transition to
a four quark plus one antiquark system, with the heavy $q\bar{q}$ pair
composed of either charm or bottom quarks.
This ansatz ultimately envisioned a simplified case wherein a spinless point particle of mass
$m_0$ interacts with coupling strength $g$ to $N$ scalar particles having masses
$m_1, m_2, \ldots, m_N$.  Pumplin then found the unintegrated light-front probability density
to have the form
\cite{Pumplin1}
\begin{align}
dP\ &=\ \frac{g^2}{(16\pi^2)^{N-1}(N-2)!}\,
    \prod_{j=1}^N dx_j\, \delta\left( 1-\sum_{j=1}^N x_j\right)
    \int_{s_0}^\infty ds\, \frac{(s-s_0)^{N-2}}{(s-m_0^2)^2}\,
    |F(s)|^2,
\label{eq:dPFock}
\end{align}
where the invariant mass is $s_0 = \sum_{j=1}^N m_j^2/x_j$, and a vertex function
$F(s)$ must be stipulated to control behavior in the ultraviolet.
In particular, if the transverse momenta and the factors of $1/x_j$ appearing
in Eq.~(\ref{eq:dPFock}) are ignored and the charm mass is taken to be much larger than
all other mass scales, one obtains the distribution prescribed by the BHPS model
\cite{BHPS} after also assuming a pointlike vertex factor $F(s) = 1$. While
these schemes are convenient and produce testable predictions for IC, a model with
more physical interactions accounting for the structure of the charm spectrum is also
desirable, and I sketch this idea in Sec.~\ref{sec:MBM} now.
%
%
%

\subsection{Meson-baryon models}
\label{sec:MBM}

As higher-mass extensions of the pion-cloud picture of nucleon structure, meson-baryon models (MBMs)
are a natural framework for studying the proton's interactions with external electromagnetic probes,
relying on the advantageous properties of time-ordered perturbative theory (TOPT) and the convolution
approach to compute corrections to the nucleon's hadronic tensor, $W^{\mu\nu}$.
\begin{figure}[h]
\centering
\includegraphics[scale=0.35]{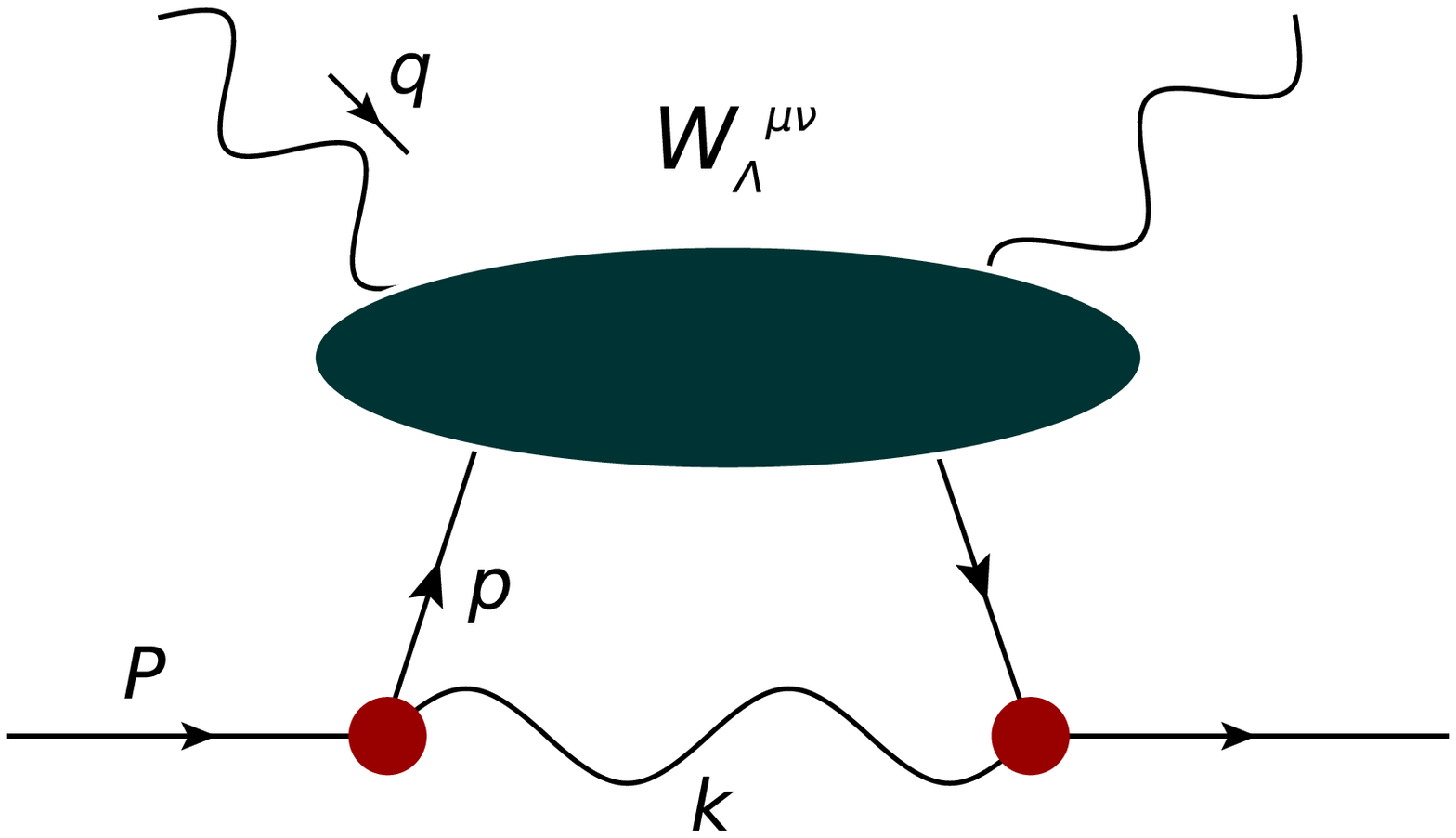} \ \ \ \ \ \ \
\includegraphics[scale=0.35]{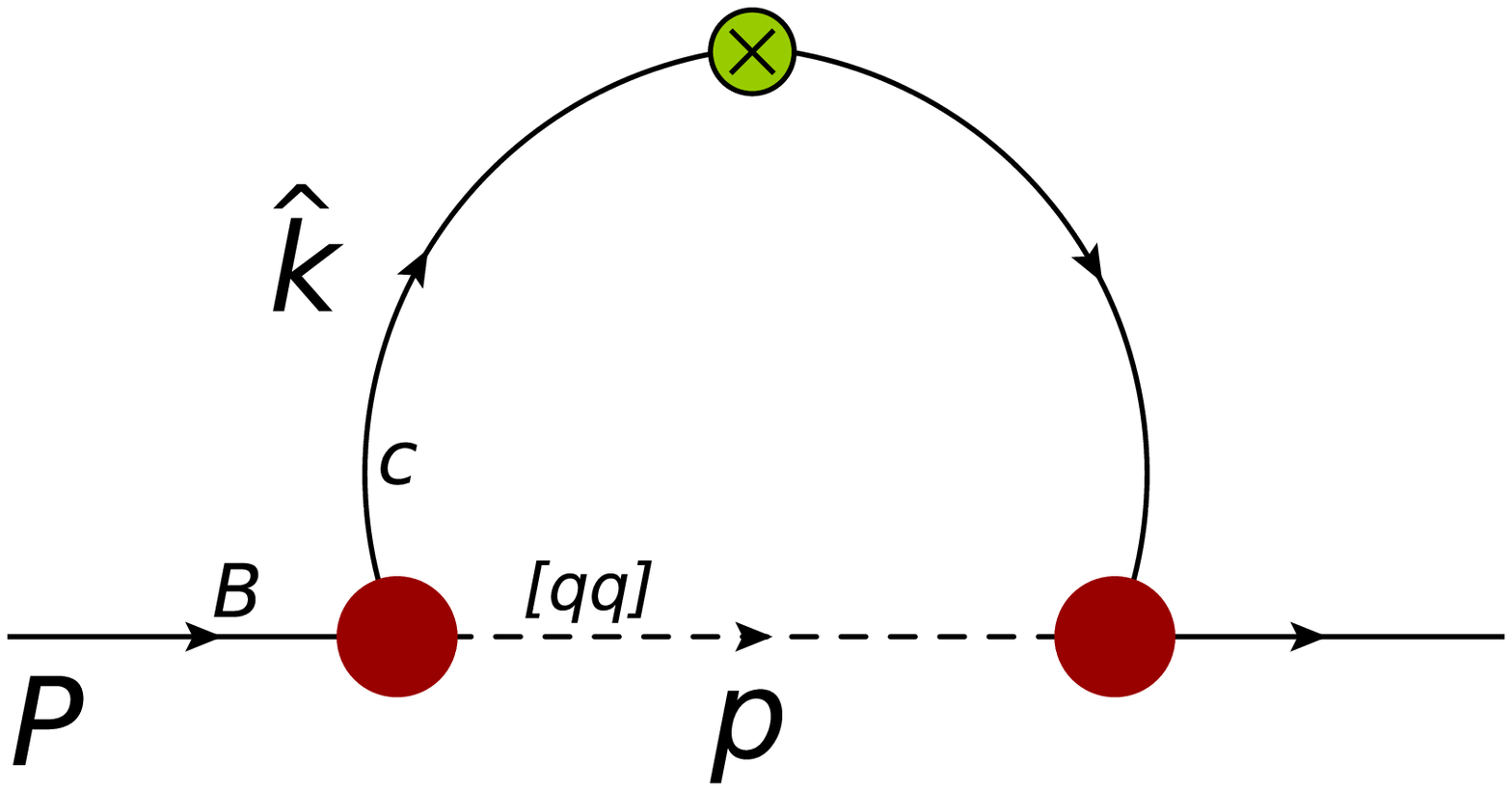}
\caption{Diagrams relevant for the dominant contribution to the charm structure function in
the MBM of Ref.~\cite{Hobbs-IC}. (Left) The TOPT diagram for the contribution of the dissociated
$\Lambda_c D^*$ state to the hadronic tensor of the proton. (Right) An analogous diagram leading
to the charm quark distribution within a spin-$1/2$ baryon (here, the $\Lambda_c$), $c_B(z)$,
assuming a quark-diquark picture for the baryon's constituent substructure; such processes are calculated
in detail in Ref.~\cite{Hobbs-IC}.
}
\label{fig:Feyn}
\end{figure}
Physically, a MBM describes the nucleon's intrinsic charm content
in a two-step ansatz formulated in terms of hadronic degrees of
freedom as well as at quark level. Moreover, unlike the BHPS
formalism with $m_c = m_{\bar{c}}$ in Sec.~\ref{sec:scalar}, MBMs can readily produce
experimentally testable asymmetries between the $c$ and $\bar c$
distributions in the nucleon.

The basic goal of these models is the probability for the nucleon to spontaneously fluctuate into
states involving an intermediate meson $M$ and baryon $B$, according to
\begin{equation}
|N\rangle\
=\ \sqrt{Z_2}\, \left| N \right.\rangle_0\
+\ \sum_{M,B} \int\! dy\, d^2\bm{k}_\perp\,
   \phi_{MB}(y,k^2_\perp)\,
   \big| M(y,\bm{k}_\perp); B(1-y,-\bm{k}_\perp) \big\rangle,
\label{eq:Fock}
\end{equation}
in which $\left| N \right.\rangle_0$ represents the undressed
three-quark nucleon state, and $Z_2$ is an associated renormalization constant.
The quantity $\phi_{MB}(y,k^2_\perp)$ is an amplitude for
the process whereby the nucleon reconfigures into an intermediate
meson $M$ carrying a fraction $y$ of the proton's longitudinal momentum
and transverse momentum $\bm{k}_\perp$, and a baryon $B$ with longitudinal
momentum fraction $1-y \equiv \bar{y}$ and transverse momentum $-\bm{k}_\perp$.
The invariant mass squared $s_{MB}$ of this intermediate state appearing in
the derivations below can then be expressed in the IMF by
\begin{equation}
s_{MB}(y,k^2_\perp)\
=\ \frac{k^2_\perp + m_M^2}{y} + \frac{k^2_\perp + M_B^2}{1-y}\ \equiv\ s\, ,
\label{eq:CoM_En}
\end{equation}
where the internal meson and baryon masses are respectively given by
$m_M$ and $M_B$.

Ultimately, the dominant mechanism determining the IC distributions
in the MBM of Ref.~\cite{Hobbs-IC} originated with the reconfiguration of the
proton into intermediate states consisting of spin-$1$ charmed mesons
$D^* = \bar{D}^{*0}$ or $D^{*-}$ and corresponding charm-containing baryons.
The associated probabilistic splitting function for this mode is related to the
amplitude of Eq.~(\ref{eq:Fock}) by \linebreak
$f_{MB}(y) = \int_0^\infty d^2\bm{k}_\perp \,|\phi_{MB}(y,k^2_\perp)|^2$,
and, due to the higher $N$-$D^*$-$\Lambda_c$ spin interaction, arises from a
linear combination of vector ($G_v$), tensor ($G_t$), and vector-tensor
interference ($G_{vt}$) pieces. {\it Viz.},
\begin{align}
f_{D^* B}(y)
&= T_B \frac{1}{16\pi^2} \int{ dk_\perp^2 \over y (1-y) }
    { |F(s)|^2 \over (s - M^2)^2 }\		
    \left[
	g^2\, G_v(y,k_\perp^2)\
     +\ {g f \over M}\, G_{vt}(y,k^2_\perp)\
     +\ \frac{f^2}{M^2}\, G_t(y,k^2_\perp)
    \right]\ ,
\label{eq:spin1-fMB}
\end{align}
where 
\begin{subequations}
\label{eq:vectABC-app}%
\begin{align}
G_v(y,k_\perp)
&= - 6 M M_B\
 +\ \frac{4(P \cdot k) (p \cdot k)}{m_D^2}\
 +\ 2 P \cdot p\ ,					\\
G_{vt}(y,k_\perp)
&= 4(M + M_B)(P \cdot p - M M_B)			\nonumber\\
& 
 -\ \frac{2}{m_D^2}
    \left[ M_B (P \cdot k)^2
	 - (M + M_B)(P \cdot k)(p \cdot k)
	 + M (p \cdot k)^2
    \right]\ ,						\\
G_t(y,k_\perp)
&= -(P \cdot p)^2\
 +\ (M + M_B)^2\, P \cdot p\
 -\ M M_B (M^2 + M_B^2 + M M_B)				\nonumber\\
& 
 +\ \frac{1}{2m_D^2}
    \Big[ (P \cdot p - M M_B) [(P-p) \cdot k]^2
	- 2 (M_B^2 P\cdot k - M^2 p\cdot k) [(P-p) \cdot k]	
\nonumber\\
&  \hspace*{1cm}
	+ 2 (P \cdot k) (p \cdot k) (2P \cdot p - M_B^2 - M^2)
    \Big]\ ,
\end{align}
\end{subequations}%
and, as depicted in the left panel of Fig.~\ref{fig:Feyn}, $p$ represents the $4$-momentum
of the interacting baryon ({\it e.g.}, $\Lambda^+_c$), $T_B$ is an isospin factor, and the
products $P \cdot p$, $P \cdot k$, $p \cdot k$ can all be evaluated in terms of explicit TOPT
expressions defined in the IMF as in the Appendices of Ref.~\cite{Hobbs-IC}. The hadronic
interaction strengths $g$ and $f$ of Eq.~(\ref{eq:spin1-fMB}) are fixed using SU$(4)$ quark model
symmetry constraints {\it \'a la} standard Lippmann-Schwinger analyses \cite{Haiden}.

A standard feature of this approach is the necessity of regulating the inevitable divergences
that appear at large $k^2_\perp$ --- a fact which follows from the essential status of MBMs as
loop corrections or dressings to the photon-nucleon vertex. Typically, regularization
is carried out phenomenologically, and implemented through a specific parametric choice for the relativistic
vertex factor $F(s)$ appearing in Eq.~(\ref{eq:spin1-fMB}), for example. The analysis of Ref.~\cite{Hobbs-IC}
made use of a Gaussian with the form
\begin{equation}
F(s) = \exp[-(s-M^2)/\Lambda^2]\ ,
\label{eq:FF_exp}
\end{equation}
in which $\Lambda$ is a cutoff parameter fixed across the various meson-baryon modes involved
in the model and determined by fitting $pp \to \Lambda_c + X$ hadroproduction data from ISR \cite{ISR};
this yielded $\Lambda = (3.0 \pm 0.2)$ GeV, leading to the predicted model bands plotted in
Fig.~\ref{fig:SFs} below.

Using this general formalism, one can compute other intermediate spin-isospin combinations for the possible
charm-containing meson-baryon states of Eq.~(\ref{eq:Fock}), and use an analogous framework
embodied by the right panel of Fig.~\ref{fig:Feyn} to determine normalized distributions of
(anti-)charm quarks within these hadronic states --- {\it e.g.}, the $c$ distribution
within the $\Lambda_c$ meson, $c_{\Lambda}(z)$, as a function of a quark momentum fraction $z$.
Having assembled these various ingredients, the MBM specifies the nonperturbative IC distribution
at the partonic threshold $Q^2 = m^2_c$ as an incoherent sum over various meson-baryon states:
\begin{subequations}
\label{eq:mesoncloud}
\begin{align}
\bar{c}(x)\
&=\ \sum_{M,B}\,
    \Big[ \int_x^1 \frac{dy}{y}\,
	  f_{MB}(y)\, \bar{c}_M\Big(\frac{x}{y}\Big)\
	+\ \int_x^1 \frac{d\bar y}{\bar y}\,
	  f_{BM}(\bar{y})\, \bar{c}_B\Big(\frac{x}{\bar{y}}\Big)
   \Big]\ ,
\label{eq:mesoncloud_cb}			\\
c(x)\
&=\ \sum_{B,M}\,
    \Big[ \int_x^1 \frac{d\bar y}{\bar y}\,
	  f_{BM}(\bar y)\, c_B\Big(\frac{x}{\bar y}\Big)\
	+\ \int_x^1 \frac{dy}{y}\,
	  f_{MB}(y)\, c_M\Big(\frac{x}{y}\Big)
    \Big]\ .
\label{eq:mesoncloud_c}
\end{align}
\end{subequations}%
\begin{figure}[h]
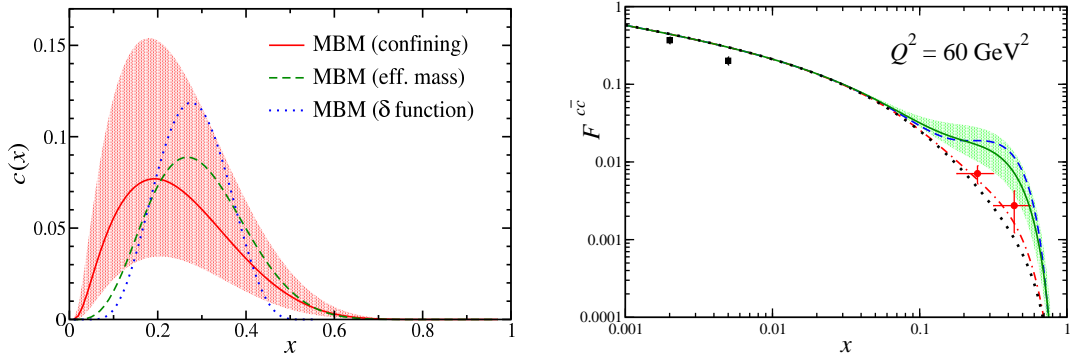

\centering
\vspace*{0.4cm}
\includegraphics[scale=0.27]{Fig-10a.eps} \ \ \ \ \ \ \
\includegraphics[scale=0.27]{IC-Q60.eps}
\caption{(Left) Intrinsic contributions to the proton's charm PDF at the starting
scale $Q^2_0 = m^2_c$ of QCD evolution under different assumed prescriptions for
the infrared behavior of the quark-hadron vertex that determines the distributions
$\bar{c}_M(z)$ and $c_B(z)$. (Right) The charm sector part of the proton structure
function $F^{c\bar{c}}_2 = \big( 4x \big/ 9 \big) \left[ c(x,Q^2) + \bar{c}(x,Q^2) \right]$
at an evolved scale of $Q^2 = 60$~GeV$^2$. The two red points belong to the highest EMC bin
$\langle Q^2 \rangle = 60$ GeV$^2$ \cite{EMC}; while these points overhang the predictions of pQCD,
they are at the lower periphery of the range predicted by the MBM informed by hadroproduction
data. 
}
\label{fig:SFs}
\end{figure}

The expressions in Eqs.~(\ref{eq:mesoncloud_cb}) and (\ref{eq:mesoncloud_c}) are the culmination
of the MBM of Ref.~\cite{Hobbs-IC} and depend crucially upon assumed schemes to counter numerical poles in TOPT energy
denominators that occur at infrared momenta in the quark-level amplitudes used to compute the
distributions $\bar{c}_M(z)$ and $c_B(z)$. As described at length in Ref.~\cite{Hobbs-IC},
three main prescriptions were employed (a ``confining'' scheme to cancel the offending TOPT denominators,
an ``effective mass'' approach in which the charm quark is taken to be sufficiently heavy as to
avoid numerical poles, and a simple delta-function assumption for the quark distributions);
after fixing UV regulators to hadroproduction data, starting-scale IC distributions such
as those shown for $c(x)$ in the left panel of Fig.~\ref{fig:SFs} were then obtained. For the sake of the
data comparisons shown in the right panel of Fig.~\ref{fig:SFs}, these distributions were evolved according
to pQCD to the empirical scale of the European Muon Collaboration (EMC) \cite{EMC}, which measured
the charm sector contribution to the $F_2(x,Q^2)$ structure function of the proton via $\mu$-Fe DIS.
%
%
%
\section{A QCD global analyses of the charm PDF}
\label{sec:GA}
The model-based treatment of IC in Sec.~\ref{sec:MBM} served the purpose of
enlightening possible physical mechanisms for generating nonperturbative charm
in a manner that led to direct constraints from experimental information. However, the ambiguity
in the resulting {\it magnitude} of IC in the proton as suggested by the panels of
Fig.~\ref{fig:SFs} demanded a more systematic numerical approach. The most comprehensive such
method involves performing a QCD global analysis of the world's data explicitly including
IC as a hypothesis. This was recently undertaken in Ref.~\cite{GA-PRL} using the $\mathcal{O}(\alpha_s)$
formalism of Hoffmann and Moore \cite{HM} for the charm structure function and, for the DGLAP starting-scale
IC itself, the parametric shapes computed in the analysis of Ref.~\cite{Hobbs-IC}. In fact, in what
follows the ``confining'' prescription corresponding to the central solid red curve in the
LHS of Fig.~\ref{fig:SFs} was used by default, and the commonly-used total momentum fraction
was taken as a proxy for the overall normalization of IC in the proton:
\begin{equation}
\langle x \rangle_{\rm IC}\ \equiv\ \int_0^1 dx\, x \left[ c(x) + \bar{c}(x) \right]\ .
\end{equation}

QCD global analyses of this quantity (as well as of an analogous fraction for intrinsic bottom
\cite{Lyonnet:2015dca}) had been carried out by a number of groups over the
years as exemplified by a recent CT14 calculation \cite{CT14} that assessed several
IC scenarios (IC$=0$, as well as BHPS and a low $x$-dominated ``sealike'' shape);
not unlike the earlier work of Ref.~\cite{Pumplin2}, this determination found
considerable levels of IC could be tolerated by a global fit --- in particular,
CT14 discovered that IC as large as $\langle x \rangle_{\rm IC} \sim 2-3\%$ could be
accommodated within their framework, depending upon the specific IC scenario assumed. 

\begin{figure}[h]
\centering
\sidecaption
\includegraphics[width=8.5cm,clip]{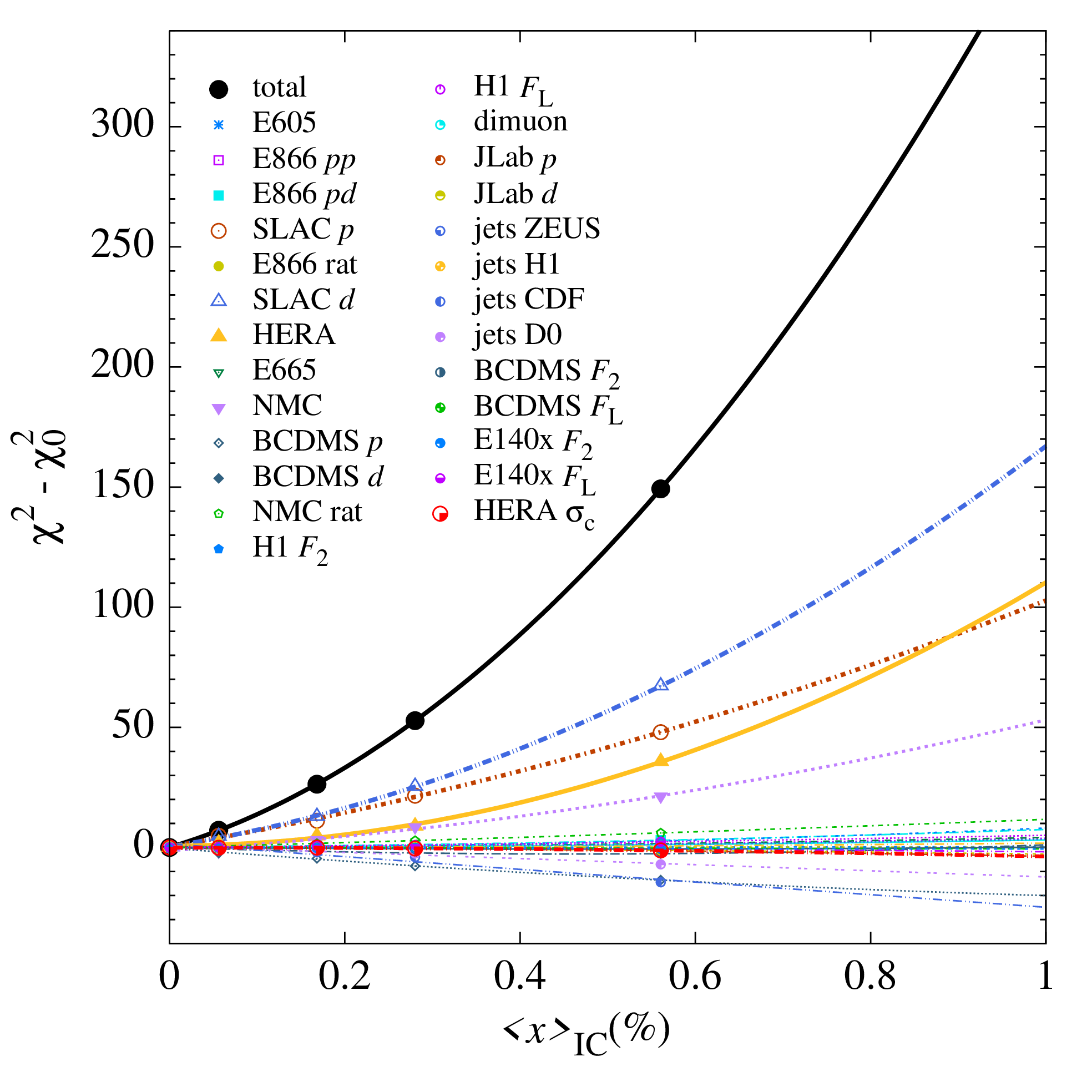}
\caption{Aside from the EMC data which were not included in the global fit shown
here, the full set of the JR14 analysis \cite{JR} was first used to study the constraints
imposed by the world's data on the total IC fraction $\langle x \rangle_{\rm IC}$. As
noted in TABLE I of Ref.~\cite{JR} and tabulated in the legend here, this
involved 4296 data points from 26 independent sets of measurements. Moreover,
the JR14 fitting technology incorporates a means of parametrizing the
effects of dynamical higher twist, target mass, and nuclear corrections, enabling substantially
less restrictive kinematical cuts: $Q^2 \ge 1$ GeV$^2$ and $W^2 \ge 3.5$ GeV$^2$.
As a result, results were additionally constrained by important inputs from fixed-target
SLAC data on proton and deuteron targets (brown circles and blue triangles,
respectively), which drove the very rapid growth in the $\chi^2$ of the global
fit relative to $\chi^2_0$ --- the corresponding value at
$\langle x \rangle_{\rm IC} = 0$.
}
\label{fig:sets}
\end{figure}

Unlike previous analyses, however, our recent effort in Ref.~\cite{GA-PRL} relied
upon an updated formalism \cite{JR} which accounted for various sub-leading $1/Q^2$ corrections
({\it e.g.}, higher-twist effects and target mass corrections) and nuclear effects for DIS observables.
This enhancement enabled a description of information at much lower $W^2$ and $Q^2$ than
typically allowed in most global fits, and we therefore leveraged this to perform our
QCD global analysis of IC with less restrictive kinematical cuts on the included data
sets ($Q^2 \ge 1$ GeV$^2$ and $W^2 \ge 3.5$ GeV$^2$).
In Fig.~\ref{fig:sets}, I summarize the contributions of the data sets included in
the global analysis of Ref.~\cite{GA-PRL} to the total growth in the $\chi^2$ resulting
from the fit. From the parabolas of Fig.~\ref{fig:sets}, it is evident that some data
sets ({\it e.g.}, the H1 $F_2$ measurements corresponding to the blue-pentagon curve
at bottom) do little to constrain $\langle x \rangle_{\rm IC}$; rather, the bulk of
the constraint to the proton's total IC (given by solid black disks) is driven by SLAC
fixed-target information --- on the deuteron (blue triangles) and proton (brown circles).
The pronounced sensitivity of these data sets to IC can be interpreted as arising indirectly
from the tightened constraints provided by the SLAC data to the light-quark sector of the
global fit, which in turn limits the flexibility of the fitting framework in tolerating
larger magnitudes for $\langle x \rangle_{\rm IC}$.

\begin{figure}[h]
\centering
\includegraphics[scale=0.34]{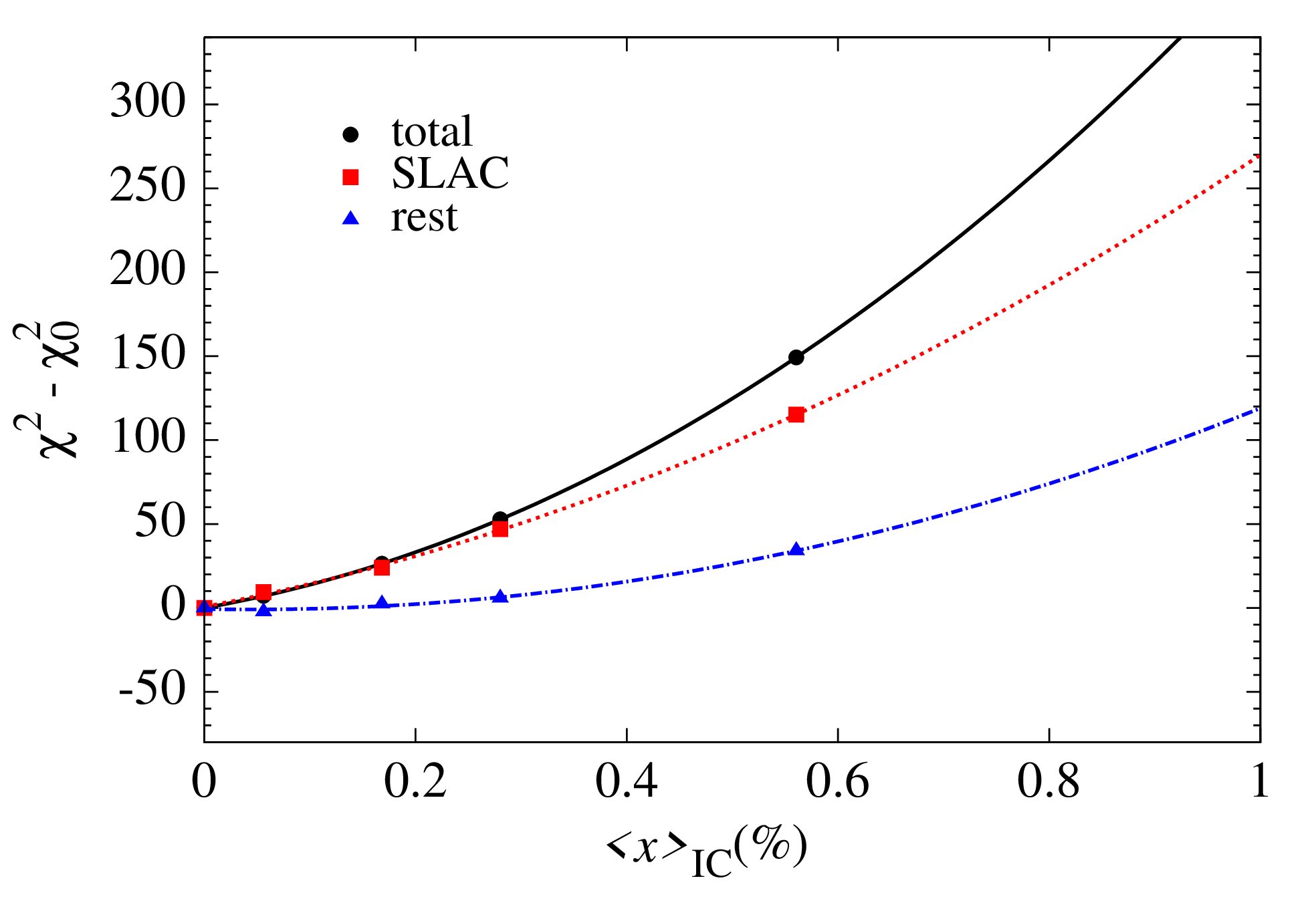} \ \ \
\includegraphics[scale=0.34]{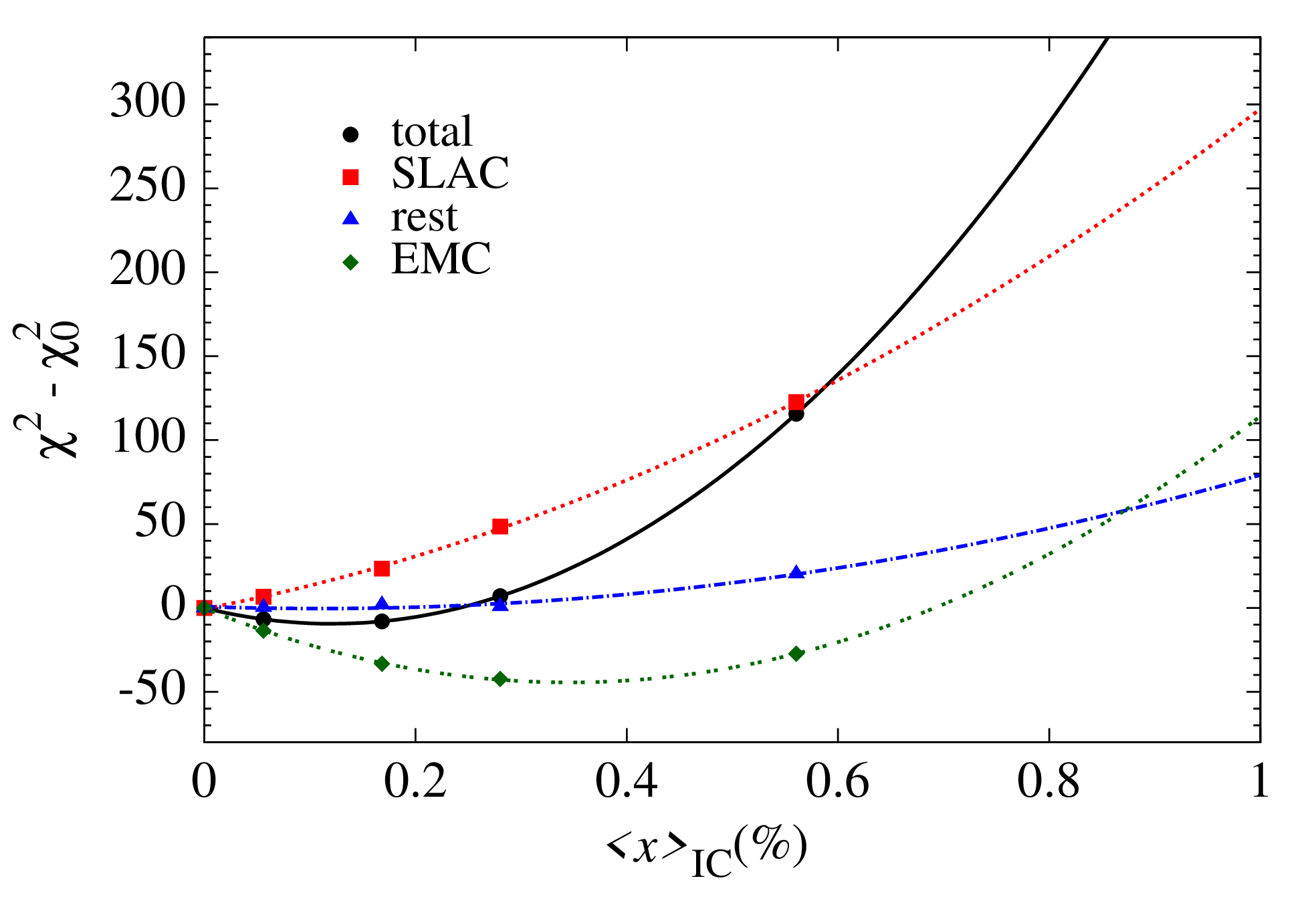}
\caption{Plots of the $\chi^2$ growth profiles in the QCD global
analysis of Ref.~\cite{GA-PRL} for runs in which EMC data were
not included (left), and for global fits constrained by EMC
(right).
}
\label{fig:GA}
\end{figure}

This point is further driven home by the left panel of Fig.~\ref{fig:GA}, which separates
the `SLAC' contributions to the $\chi^2$ profile from the `rest' of the data set summarized in
Fig.~\ref{fig:sets}. In the end, this global fit places quite stringent constraints on
the magnitude of IC: $\langle x \rangle_{\rm IC} < 0.1\%$ at the $5\sigma$ level.

An additional motivation for the analysis of Ref.~\cite{GA-PRL}, however, was the need
to evaluate the suggestive measurements of EMC \cite{EMC}. As has been pointed out numerous times
in the literature, the highest $Q^2$ bins of the EMC data seem to provide some hint of the
existence of IC, but before Ref.~\cite{GA-PRL} this information had not been treated in the
context of a global analysis. Ultimately, incorporating the EMC points into the data set
summarized in Fig.~\ref{fig:sets} led to a slight preference for nonzero IC, with the
EMC set alone favoring $\langle x \rangle_{\rm IC} = 0.3-0.4\%$ as indicated by the
minimum in the green double-dotted curve on the RHS of Fig.~\ref{fig:GA}. The modest
preference for nonzero IC is diluted by the presence of other mitigating data sets,
however, and the full global fit results in $\langle x \rangle_{\rm IC} = 0.13 \pm 0.04\%$
--- a significantly reduced magnitude relative to the results of earlier analyses like
Ref.~\cite{CT14}. We also found the EMC points to be poorly fit (with $\chi^2/{\rm datum} = 4.3$)
and in some apparent tension with lower $Q^2$ points from HERA \cite{HERA}. These issues necessitate further
evaluation of the EMC set, and demand additional experimental measurement which might clarify
the magnitude of IC without residual tension or ambiguity.

%
%
\section{Recent developments in IC phenomenology}
\label{sec:recent}
Following the completion of the work in Ref.~\cite{GA-PRL}, a number of intriguing developments
in the phenomenology of IC have emerged, of which I highlight a couple of recent examples.

Partly motivated by the work of Refs.~\cite{GA-PRL,Reply}, Ball {\it et al}.~of the NNPDF
Collaboration announced an independent global analysis \cite{NNPDF} of the charm PDF
in which the possibility of intrinsic charm was explicitly allowed via separate ``Perturbative''
and ``Fitted'' distributions which were then constrained through the neural networks-based
methodology of NNPDF. Moreover, like Refs.~\cite{GA-PRL,Reply} this work also incorporated the
provocative EMC data, confronting them with a global fitting technology that does not presuppose
a particular model-derived shape for the IC distribution; the relaxation of this aspect
of typical QCD global fits affords the NNPDF framework greater flexibility in describing the
world's data, but can also result in distributions that are difficult to reconcile with the
predictions of model-building. This fact is evident in the especially hard shape for
the ``Fitted'' charm distribution obtained in Fig.~3 of Ref.~\cite{NNPDF}, which corresponded to
$\langle x \rangle_{\rm IC} = 0.7 \pm 0.3\%$, and demands further study.

Aside from direct measurements of the charm structure function, more oblique means of accessing
the nucleon's nonperturbative charm content can be constructed. An archetypal example of this
sort of approach can be found in the precise observation of prompt neutrino production
\cite{Hallsie} in dedicated cosmic-ray experiments like IceCube \cite{Laha}. The analysis
in this latter reference demonstrates that IC in accord with the upper reaches of the model predictions of Ref.~\cite{Hobbs-IC}
(which allows larger IC normalizations and preceded the more systematic global analysis of
Ref.~\cite{GA-PRL}) might in principle be observable in the IceCube prompt neutrino spectrum;
this suggest that such measurements could serve the role of an additional avenue to either
observing or constraining the proton's nonperturbative charm.
%
%
\section{Conclusion and outlook}
\label{sec:conc}
I have described the results of several closely related lines of investigation into the intrinsic
charm problem. Given that direct experimental information is relatively limited, much of this work has
been constrained in its ability to recommend a robust value for, {\it e.g.}, the total IC momentum
fraction $\langle x \rangle_{\rm IC}$. The present situation therefore seems to be one in which
the reach of modeling and analyses of data is approaching a point of diminishing return, and additional
empirical inputs would be invaluable. For this purpose modern, direct measurements of $F^{c\bar{c}}_2(x,Q^2)$
--- especially at large $x$ and low/intermediate $Q^2$ along the lines of the original EMC data --- could
better constrain both models and global analyses and settle some of the questions related to the
interpretation of the EMC set itself as raised in Refs.~\cite{GA-PRL,Reply}. Measurements
of this type might ideally be carried out at a future electron-ion collider \cite{EIC} and be complemented
by work at the proposed AFTER@CERN fixed-target $pp$ experiment \cite{AFTER}, which would putatively operate at
$\sqrt{s} = 115$ GeV; this might similarly be the case for a suggested measurement of the forward production
of $Z$ bosons in coincidence with charm-containing jets at LHCb \cite{Williams}, possibly exposing
the behavior of the charm PDF in the critical valence region.
%
%
\section{Acknowledgements}
\label{sec:Ack}
I thank Wally Melnitchouk, Tim Londergan, and Pedro Jimenez-Delgado for discussions and collaboration
on various portions of this work. For additional useful conversations, I thank Stan Brodsky, Ranjan Laha,
Gerald Miller, and Jean-Phillipe Lansberg. This work was supported by the U.S.~Department of Energy Office of
Science, Office of Basic Energy Sciences program under Award Number DE-FG02-97ER-41014.
%

\begin{thebibliography}{}
%
\bibitem{BHPS}
S.J. Brodsky, P.~Hoyer, C.~Peterson, N.~Sakai, Phys. Lett. \textbf{B93}, 451
  (1980)

\bibitem{Chang-Yan}
S.J. Chang, R.G. Root, T.M. Yan, Phys. Rev. \textbf{D7}, 1133 (1973)

\bibitem{Lepage}
G.P. Lepage, S.J. Brodsky, Phys. Rev. \textbf{D22}, 2157 (1980)

\bibitem{Pumplin1}
J.~Pumplin, Phys. Rev. \textbf{D73}, 114015 (2006), \texttt{hep-ph/0508184}

\bibitem{Pumplin2}
J.~Pumplin, H.L. Lai, W.K. Tung, Phys. Rev. \textbf{D75}, 054029 (2007),
  \texttt{hep-ph/0701220}

\bibitem{Hobbs-IC}
T.J. Hobbs, J.T. Londergan, W.~Melnitchouk, Phys. Rev. \textbf{D89}, 074008
  (2014), \texttt{1311.1578}

\bibitem{Haiden}
J.~Haidenbauer, G.~Krein, U.G. Meissner, L.~Tolos, Eur. Phys. J. \textbf{A47},
  18 (2011), \texttt{1008.3794}

\bibitem{ISR}
P.~Chauvat et~al. (R608), Phys. Lett. \textbf{B199}, 304 (1987)

\bibitem{EMC}
J.J. Aubert et~al. (European Muon), Nucl. Phys. \textbf{B213}, 31 (1983)

\bibitem{GA-PRL}
P.~Jimenez-Delgado, T.J. Hobbs, J.T. Londergan, W.~Melnitchouk, Phys. Rev.
  Lett. \textbf{114}, 082002 (2015), \texttt{1408.1708}

\bibitem{HM}
E.~Hoffmann, R.~Moore, Z. Phys. \textbf{C20}, 71 (1983)

\bibitem{Lyonnet:2015dca}
F.~Lyonnet, A.~Kusina, T.~Ježo, K.~Kovarík, F.~Olness, I.~Schienbein, J.Y.
  Yu, JHEP \textbf{07}, 141 (2015), \texttt{1504.05156}

\bibitem{CT14}
S.~Dulat, T.J. Hou, J.~Gao, J.~Huston, J.~Pumplin, C.~Schmidt, D.~Stump, C.P.
  Yuan, Phys. Rev. \textbf{D89}, 073004 (2014), \texttt{1309.0025}

\bibitem{JR}
P.~Jimenez-Delgado, E.~Reya, Phys. Rev. \textbf{D89}, 074049 (2014),
  \texttt{1403.1852}

\bibitem{HERA}
H.~Abramowicz et~al. (ZEUS, H1), Eur. Phys. J. \textbf{C73}, 2311 (2013),
  \texttt{1211.1182}

\bibitem{Reply}
P.~Jimenez-Delgado, T.J. Hobbs, J.T. Londergan, W.~Melnitchouk, Phys. Rev.
  Lett. \textbf{116}, 019102 (2016), \texttt{1504.06304}

\bibitem{NNPDF}
R.D. Ball, V.~Bertone, M.~Bonvini, S.~Carrazza, S.~Forte, A.~Guffanti, N.P.
  Hartland, J.~Rojo, L.~Rottoli (NNPDF) (2016), \texttt{1605.06515}

\bibitem{Hallsie}
R.~Enberg, M.H. Reno, I.~Sarcevic, Phys. Rev. \textbf{D78}, 043005 (2008),
  \texttt{0806.0418}

\bibitem{Laha}
R.~Laha, S.J. Brodsky (2016), \texttt{1607.08240}

\bibitem{EIC}
A.~Accardi et~al., Eur. Phys. J. \textbf{A52}, 268 (2016), \texttt{1212.1701}

\bibitem{AFTER}
S.J. Brodsky, A.~Kusina, F.~Lyonnet, I.~Schienbein, H.~Spiesberger, R.~Vogt,
  Adv. High Energy Phys. \textbf{2015}, 231547 (2015), \texttt{1504.06287}

\bibitem{Williams}
T.~Boettcher, P.~Ilten, M.~Williams, Phys. Rev. \textbf{D93}, 074008 (2016),
  \texttt{1512.06666}
%
\end{thebibliography}
%
%

%
\end{document}